\documentclass[aas2pp4,natbib209]{emulateapj}

\newcommand\icarus{\ref@jnl{Icarus}}%

\usepackage{epstopdf}

\slugcomment{2016 Jan 27}

\begin{document}


\title{Mapping CO Gas in the GG Tauri A Triple System with 50 AU Spatial Resolution}

\author{Ya-Wen Tang$^1$, Anne Dutrey$^2$, St\'ephane Guilloteau$^2$, Edwige Chapillon$^3$,
Vincent Pietu$^3$, Emmanuel Di~Folco$^2$,  Jeff Bary$^4$, Tracy Beck$^5$, Herv\'e Beust$^6$, Yann Boehler$^7$,
Fr\'ederic Gueth$^3$, Jean-Marc Hur\'e$^2$, Arnaud Pierens$^2$,
Michal Simon$^8$, }
\affil{$^{1}$Academia Sinica, Institute of Astronomy and Astrophysics, Taipei, Taiwan}
\affil{$^{2}$Universit\'e de Bordeaux, Observatoire Aquitain des Sciences de l'Univers,
CNRS, UMR 5804, Laboratoire d'Astrophysique de Bordeaux,
2 rue de l'Observatoire, BP 89, F-33271 Floirac Cedex, France}
\affil{$^{3}$IRAM, 300 rue de la piscine, F-38406 Saint Martin d'H\`eres Cedex, France}
\affil{$^{4}$Department of Physics and Astronomy, Colgate University, 13 Oak Drive, Hamilton, NY 13346, USA}
\affil{$^{5}$Space Telescope Science Institute,3700 san Martin Dr. Baltimore, MD 21218, USA}
\affil{$^{6}$Universit\'e de Grenoble, IPAG, Saint Martin d'H\`eres, France}
\affil{$^{7}$Centro de Radioastronom\`ia y Astrof\`isica, UNAM, Apartado Postal 3-72, 58089 Morelia, Michoac\`an, Mexico}
\affil{$^{8}$Stony Brook University, Stony Brook, NY 11794-3800, USA}

\email{ywtang@asiaa.sinica.edu.tw}

\begin{abstract}
We aim to unveil the observational imprint of physical mechanisms
that govern planetary formation in the young, multiple system GG Tau A.
We present ALMA observations of $^{12}$CO and  $^{13}$CO 3-2 and 0.9 mm continuum emission with 0.35" resolution.
The $^{12}$CO 3-2 emission, found within the cavity of the circumternary dust ring (at radius $< 180$ AU) where no $^{13}$CO emission is detected,
confirms the presence of CO gas near the circumstellar disk of GG Tau Aa.  The outer disk and
 the recently detected hot spot lying at the outer edge of the dust ring are mapped both in $^{12}$CO and $^{13}$CO. The gas
emission in the outer disk can be radially decomposed as a series of slightly overlapping Gaussian rings,
suggesting the presence of unresolved gaps or dips. The dip closest to the disk center lies at a radius very
close to the hot spot location at $\sim250-260$~AU. The CO excitation conditions
indicate that the outer disk remains in the shadow of the ring. The hot spot probably results from local
heating processes. The two latter points reinforce the hypothesis that the hot spot is created by
an embedded proto-planet shepherding the outer disk.
\end{abstract}


\section{Introduction}

The discovery of numerous exoplanets in binary and higher order multiple systems, whether orbiting one of the stellar companions
or both, clearly demonstrates that planets form and remain in stable orbits in these dynamically complex systems
\citep[e.g. the Kepler systems, 16, 34, and 35;][]{doyle11, welsh12}.  Given that roughly 50\% of Sun-like star are in
 in binary or higher-order multiple systems \citep{duquennoy91,raghavan2010}, a substantial fraction of planets
 in the Galaxy have formed in such systems.  A complete understanding
of planet formation requires us to understand the environments and mechanisms that give rise to such planetary systems.

The massive circum-ternary Keplerian disk surrounding GG Tau Aa, Ab1 and Ab2 \citep{difolco14} consists of a
dense gas and dust annulus or ring that extends from $\sim$190 to 280 AU from the central stars and a lower density outer region
that extends out to $\sim$800~AU \citep[e.g.][]{dutrey94,guilloteau99,pietu11}.  The total mass of the disk,
$\sim$0.15~M$_\odot$ \citep{dutrey94}, makes it one of the most massive disks orbiting a low-mass stellar system.
Interestingly, 80\% of the disk mass is confined within the narrow ring despite accounting for $<$~8\% of the
disk's volume.  The outer disk, which is hidden from the stellar light by the dense ring, is very cold.  \citet{dutrey14} measured
a dust temperature of $\sim$~8~K at 300~AU and \citet{guilloteau99} derived a kinetic temperature of $\sim$ 20~K at the same
distance from the stars.  \citet{dutrey14} also noted that the dust ring appears smooth and homogeneous at 0.5 millimeter (mm) emission.

In the near-infrared (near-IR), the ring is optically thick. Detailed models of
scattered light emission at wavelengths between 0.8 and 3.6~$\mu$m found that the
emission short ward of 2~$\mu$m originates from 50~AU above and below the disk
mid-plane at the inner edge of the ring \citep{duchene04}.  This value
is consistent with the thickness of
$\sim$~120~AU derived from the position offset between the ring detected at near-IR
and mm wavelengths \citep{guilloteau99}.  The dust ring is very homogeneous with the
exception of a dark lane that is visible on the western side, perhaps the result of a shadow
cast by dusty material located inside the inner truncation radius of the disk \citep{itoh14}.

Evidence for gas flowing from the outer ring towards the central stars
has been inferred from $^{12}$CO 2-1 gas emission \citep{guilloteau01} and has been
strongly suggested by a high-spatial resolution map of near-IR H$_2$ gas
confined within the inner cavity \citep{beck12}.  Scattered light images also
hinted at the presence of optically thin dust located between the stars and the
dust ring \citep[e.g.][]{roddier96,silber00}.  A recent ALMA study of $^{12}$CO~6-5
and IRAM $^{12}$CO~2-1 data clearly shows several fragments of CO gas with
a kinematic signature indicating that they were falling from the ring toward the
inner stellar companions, likely feeding the circumstellar disk associated with the
primary star \citep{dutrey14}.  In this same study, a puzzling CO ``hot spot" was
detected at the outer edge of the dense ring at a radius of $\simeq$~260~AU.  With a temperature estimated to be $>$~40~K, the spot is twice as warm as the
surrounding gas.  Given the location of this hot spot at the edge of the unusually dense ring,
one exciting, although highly speculative, interpretation of this structure is that it
represents an embedded companion still in the process of forming.

In this paper, we report our Cycle 1 ALMA observational results of
thermal dust at 0.9\,mm and line emission from $^{12}$CO 3-2 and
$^{13}$CO 3-2 at 0$\farcs$35 angular resolution.
We describe the radial profile of the CO gas
distribution in the outer ring and the extended region of the disk as well
as the CO emission centered on the hot spot.
The observations and data reduction are described in section 2.
Section 3 presents the analysis of the CO outer disk and is followed by
a discussion in Section 4.

%
\begin{figure*}
\centering
\centering
\includegraphics[scale=0.7]{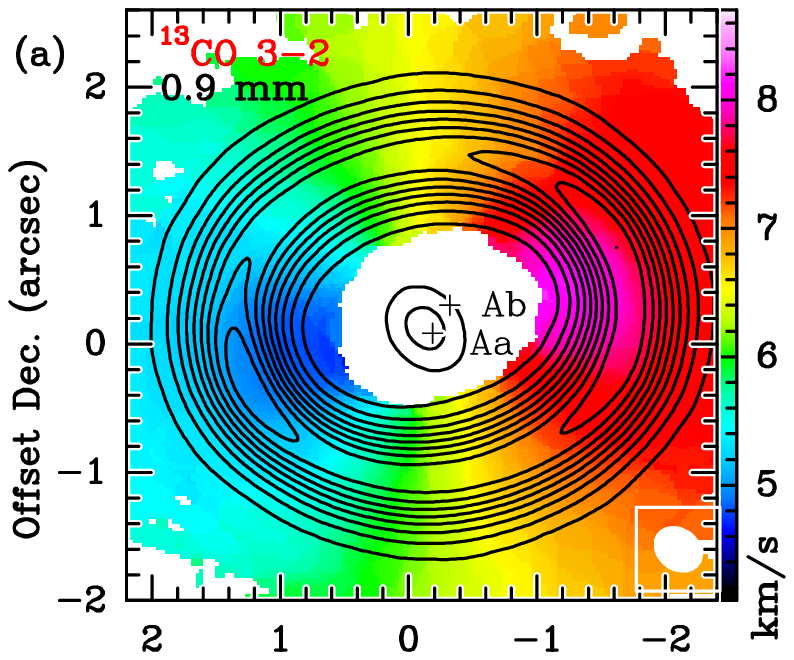}
\includegraphics[scale=0.7]{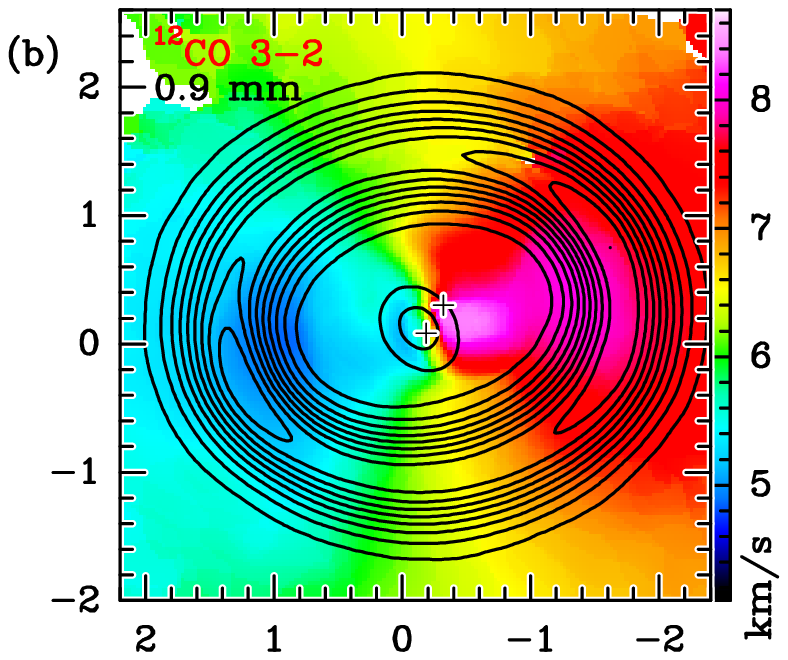}
\includegraphics[scale=0.7]{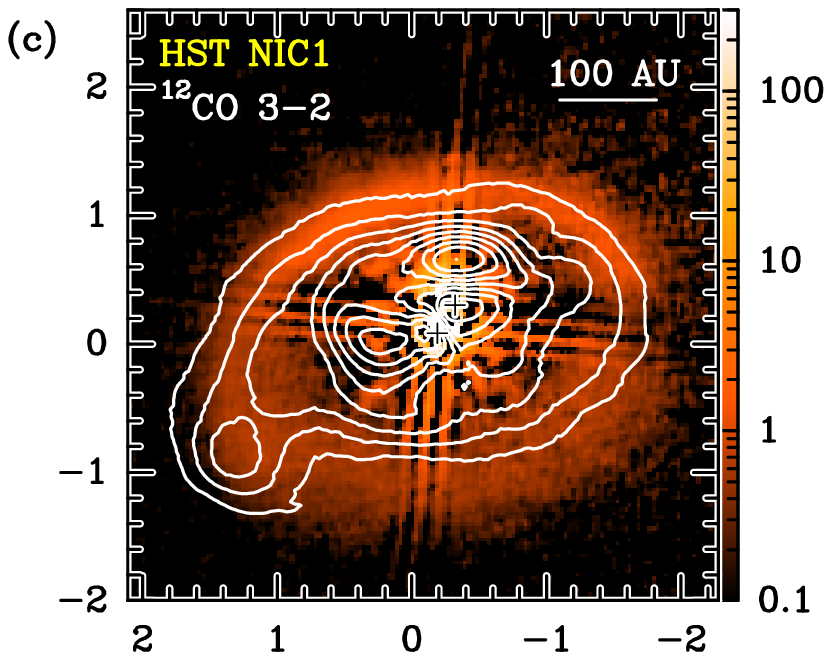}
\includegraphics[scale=0.7]{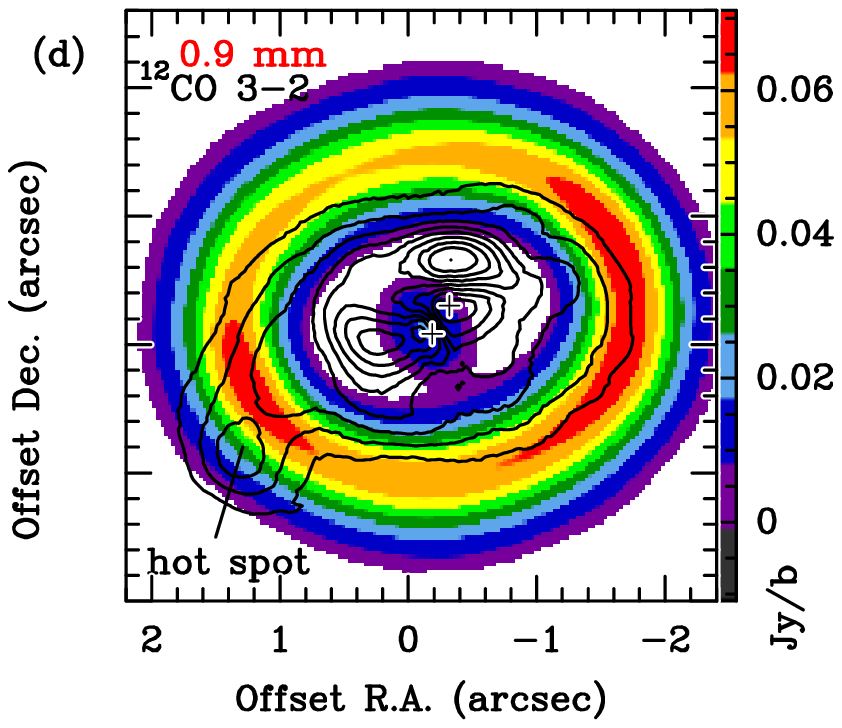}
\includegraphics[scale=0.7]{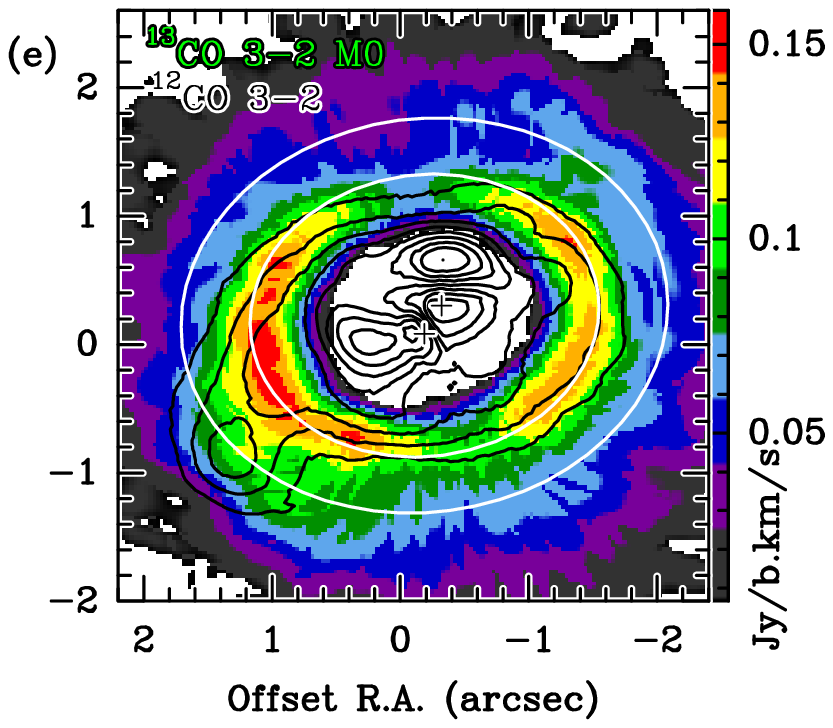}
\includegraphics[scale=0.7]{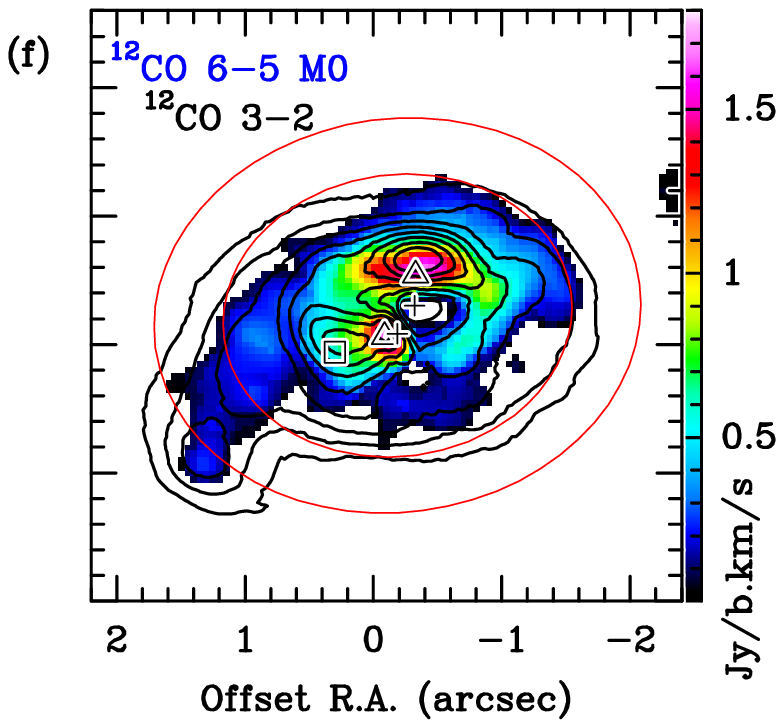}
\caption{Panel a and b: Continuum emission map at 0.9 mm (contours) overlaid on the intensity weighted velocity map
(color scale) of $^{13}$CO 3-2 in panel a and of $^{12}$CO 3-2 in panel b. Contours are 10, 20, 30, ..., 80, 90\% of
the peak intensity 42.8 mJy/beam. Panel c: $^{12}$CO 3-2 moment 0 map overlaid on the HST polarization image from
\citet{silber00} in color scale. Panel d: Moment 0 map of $^{12}$CO 3-2 overlaid on the 0.9 mm continuum (color scale).
Panel e: $^{12}$CO (contours) moment 0 map overlaid on the $^{13}$CO 3-2 moment 0 map (color scale). Panel f: $^{12}$CO
(contours) moment 0 map overlaid on the $^{12}$CO 6-5 moment 0 map (color scale). The triangles and the square mark the
peak locations of CO 6-5 and 2-1, respectively. The contours in panel c, d, e, f are 20, 30, 40,..., 80, 90\% of the
peak intensity 0.85 Jy/beam km/s.  Note that the $^{12}$CO 3-2 emission is not shown in its full extension.
Pluses mark the locations of Aa and Ab stars. The ellipses are the inner and outer radii of the dust ring at
191 AU and at 266 AU, respectively.}
\label{fig:gas_maps}
\end{figure*}

\begin{figure*}
\centering
\includegraphics{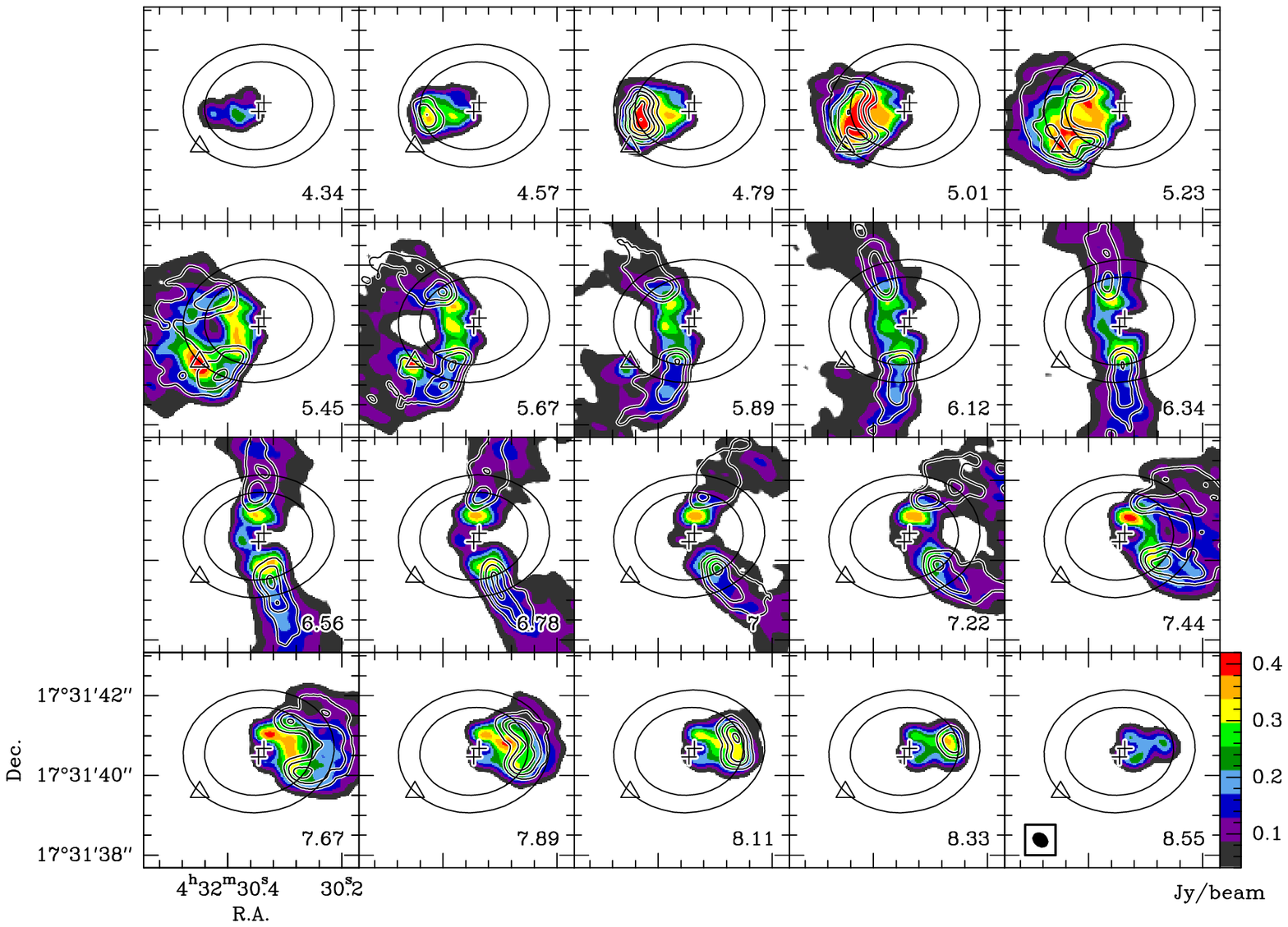}
\caption{Channel maps of $^{12}$CO 3-2 (color scale) and $^{13}$CO 3-2 (contours). Contours are 0.2,0.4,0.6,0.8 and 1
of the peak intensity 0.30 Jy/beam. Pluses and ellipses are as in Figure \ref{fig:gas_maps}, but the triangle
marks here the location of the hot spot. The number at bottom of each
panel is the velocity with respect to the standard of rest (V$_{\rm LSR}$) in km/s. The angular resolution is shown as
a black ellipse in the panel of V$_{\rm LSR}$ 8.55 km/s.}
\label{fig:channels}
\end{figure*}

\section{Observations and Results}

\subsection{Observations}
GG Tau A was observed with the ALMA Band~7 receiver during Early Science (Cycle 1, project "2012.1.00129.S").
Observations were made in three execution blocks on November 18 and 19, 2013.
The array included 29 antennas with projected baselines ranging from 15\,m to 1282\,m.
The frequency setup covered the lines of $^{12}$CO and $^{13}$CO 3-2 at 345.8 and 330.6 GHz
with two spectral windows (SPWs), each providing 1920 channels of 35.3 kHz effective resolution
($\sim 0.031$\,km/s) and 58.6 MHz bandwidth. The other two SPWs were
set to detect the continuum emission, and each SPW has 3840 channels within a bandwidth of 1.875 GHz, and the spectral resolution is 977 kHz.
The calibration (bandpass, phase, amplitude, flux) was performed using CASA\footnote{http://casa.nrao.edu}
v4.1. The bandpass calibration was done using J0423-0120 (aka J0423-013, flux $\sim$ 1.4Jy at 343.5GHz),
which also provided the flux scale. The phase and amplitude calibrator is J0509+1806, located  $\sim 9^\circ$ away from the source.

\subsection{Results}
CO and continuum maps were generated from the visibility data using the
GILDAS\footnote{https://www.iram.fr/IRAMFR/GILDAS} package. A high dynamic range was obtained by self-calibrating the
continuum data and applying this solution to the line data. In general, the results and maps presented in
this paper use a robust weighting with the exception of the spectra presented in Sec.\ref{ana:co_dyn}.
Robust weighting 
gives better angular resolution as follows: 0$\farcs$41$\times$0$\farcs$33 with position angle (P.A.)
of 52$\degr$, 0$\farcs$43$\times$0$\farcs$34 with P.A. of 51.3$\degr$, and 0$\farcs$39$\times$0$\farcs$35 with P.A. of
80$\degr$ for the $^{12}$CO 3-2, $^{13}$CO 3-2 and 0.9 mm continuum, respectively.  The achieved 1-$\sigma$ sensitivity
is 9.6 mJy/beam (0.9~K) per velocity channel (35.3 kHz) for the $^{12}$CO and $^{13}$CO line measurements, respectively.
The achieved sensitivity for the 0.9~mm continuum observations is 0.2~mJy/beam.  All results have been corrected for
the source proper motion (17, -19) mas~yr$^{-1}$ \citep{ducourant05,dutrey14} and precessed to Epoch J2000.

Continuum emission from dust in the GG Tau A system is clearly detected at 0.9~mm and displays characteristics similar to
previous continuum observations at 0.45, 1.3, and 3~mm including an unresolved emission peak at the location of the primary
companion (GG Tau Aa) and its circumstellar disk as well as a narrow elliptical ring of emission corresponding to the dense,
dusty circum-ternary ring (see contours in Figure \ref{fig:gas_maps} panels a and b).  The best-fit parameters for the inner and
outer radius, position angle, and disk inclination are 1$\farcs$36 (191 AU), 1$\farcs$90 (266 AU),
7.5$\degr$ and $36.4\degr$, respectively, with negligible formal errors at this precision.

Maps of $^{12}$CO and $^{13}$CO 3-2 line emission are presented in Figure~\ref{fig:gas_maps}.  The $^{13}$CO 3-2 image
reveals a cavity within the dusty ring in the vicinity of the stellar companions, a result consistent with the dust continuum emission
and previous $^{13}$CO 1-0 and 2-1 results \citep{dutrey94,guilloteau99}.  The intensity weighted velocity map of the $^{13}$CO 3-2
emission in Figure~\ref{fig:gas_maps}a shows that this line mostly traces the Keplerian motion of gas residing in the circum-ternary disk.
By contrast, $^{12}$CO 3-2 emission extends inside the cavity down to the positions of the stellar companions (see Figure~\ref{fig:gas_maps}).
The peak $^{12}$CO 3-2 emission within the cavity is physically located between those associated with $^{12}$CO 6-5 tracing warm gas
near the stellar companions and $^{12}$CO 2-1 tracing colder gas near the inner edge of the ring.  Such a stratification is typical of a
gradient in the excitation and physical conditions (i.e., temperature and density) of the gas in this environment as noted by \citet{dutrey14}.
In the $^{12}$CO 3-2 line emission contours in Figure~\ref{fig:gas_maps}c-f, a long extension of gas reaches toward the southeast terminating
near the outer edge of the dense circum-ternary ring some 260~AU from the disk center and spatially coincident with the location of the ``hot
spot'' previously detected in the CO 6-5 mentioned above \citep{dutrey14}.  The hot spot is also visible in the $^{12}$CO 3-2 channel maps in
Figure~\ref{fig:channels}, at velocities between 5.23 and 6.12 km~s$^{-1}$.

\begin{figure}
\includegraphics[width=0.5cm,angle=0.0]{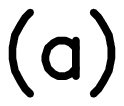}
\includegraphics[width=8.0cm,angle=0.0]{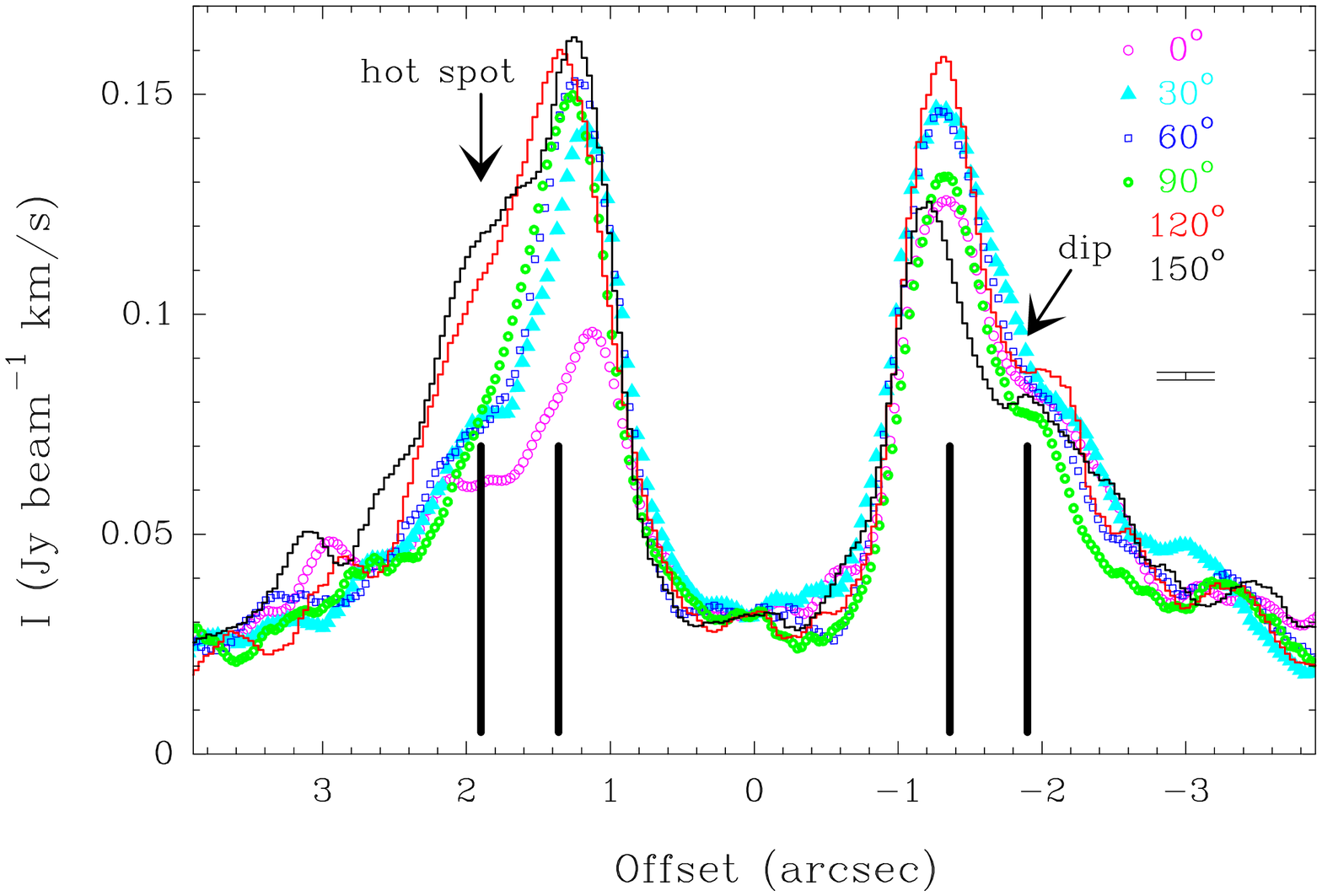}
\vspace{0.1cm}
\\
\includegraphics[width=0.5cm,angle=0.0]{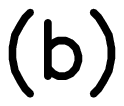}
\includegraphics[width=8.0cm]{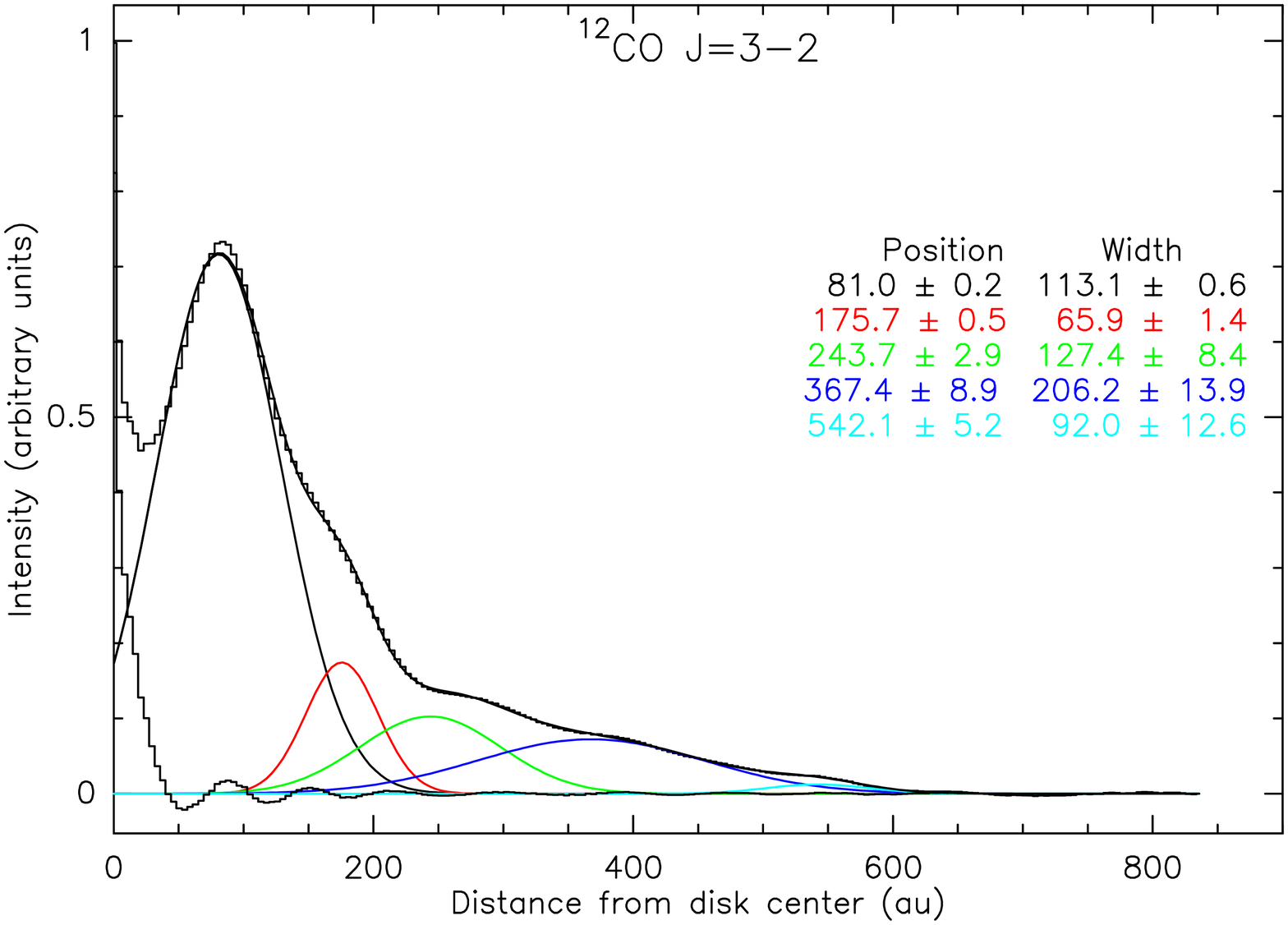}
\vspace{0.1cm}
\\
\includegraphics[width=0.5cm,angle=0.0]{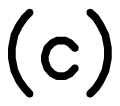}
\includegraphics[width=8.0cm]{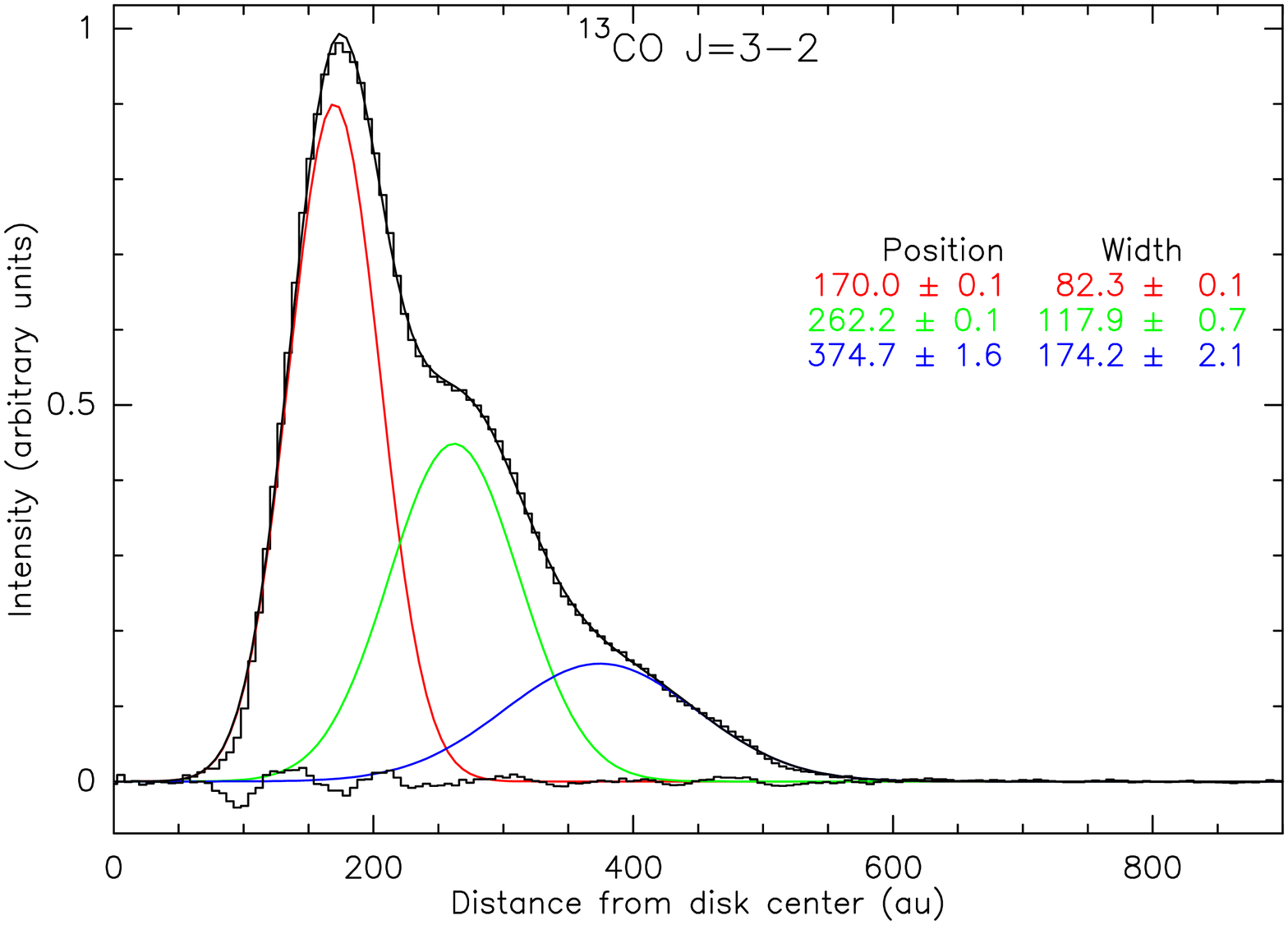}
\caption{(a) Top panel: Plots of the $^{13}$CO 3-2 intensity versus offset from the disk center at various azimuth angle $\theta$.
Positive offsets correspond to the East of the disk.
The location of the hot spot feature (down arrow) is at $\theta$ of 123$\degr$ at offset 1$\farcs$90 (266 AU). The vertical lines
mark the inner radius at 1$\farcs$36 (191 AU) and the outer radius at 1$\farcs$9 (266 AU) of the dust ring.
(b) middle and (c) lower panels: Radial average (over the western half of the disk) of the $^{12}$CO and $^{13}$CO 3-2
moment 0 maps, their decomposition in Gaussians, and the residuals.
The radial position and FWHM of the Gaussians (in AU) are indicated with their formal errors in}
their respective colors.
All the plots are produced with de-projected maps.
\label{fig:cocut}
\end{figure}

\section{Analysis of the CO radial distribution}
\label{sec:analysis}
\subsection{CO radial distributions}
\label{sec:ana:radial}

In Figure~\ref{fig:cocut}a, six radial profiles of the de-projected integrated flux or moment 0 map of the
$^{13}$CO 3-2 line are displayed.  The disk rotation axis, which is 7.5$\degr$ East of 
North, is used as the azimuthal reference or $\theta=0\degr$ for the radial cuts.  All radial profiles exhibit a
sharp inner edge, and extend slightly inside the inner truncation radius for the dusty ring.  Moving radially
outward, the profiles do not decrease monotonically beyond the inner edge of the ring.  The $^{13}$CO 3-2
emission is clearly azimuthally asymmetric with a notable bright spot located at an offset of
$\sim$1$\farcs$9 and 120~$\le$~$\theta$~$\le$~150~$\degr$, a position that is coincident with the hot spot
feature.

In Figure~\ref{fig:cocut}b and c, we produce the average radial profile over the western
side of the disks from the moment 0 maps of both $^{12}$CO and $^{13}$CO 3-2, avoiding contributions
from the feature on the eastern side.  The radial profile
for the western half of the disk is decomposed into a series of partially overlapping Gaussians for both CO lines.
The positions of the peaks (distance from center in AU) and widths (FWHM in AU) of these Gaussians and their
corresponding formal errors are presented in Figure~\ref{fig:cocut}b and c.  In $^{12}$CO, the Gaussian with the
brightest amplitude peaks at 81 au, coincident with the gas streamer described in \citet{dutrey14}.  At radii $\le$30~AU,
the residuals likely are dominated by the CO disk surrounding the primary, Aa.  In the dusty ring and extended disk,
the $^{12}$CO and $^{13}$CO radial distributions can be decomposed into three bright Gaussians centered
around 170-176~AU, 240-260~AU and 370~AU.  A fourth and outermost Gaussian is observed at a radius of 540~AU,
only in $^{12}$CO.  Given the widths of these Gaussian features and a spatial resolution of 50~AU, all are spatially
resolved with the exception of the one that peaks near 170~AU.  This feature also happens to be located just slightly
inside the inner truncation radius for the dusty ring 190~AU.
The difference in apparent inner radii  is likely due to the high contrast in opacity between
the $^{13}$CO 3-2 and 0.9\,mm dust emissions, which
is on the order of 3500 assuming standard dust properties
\citep{beckwith+90}, a local line-width of 0.3 km/s,
a temperature of 20 K and a $^{13}$CO abundance of 1.4$\times$10$^{-6}$
\citep[see][for details]{dutrey96}. A very small amount of dust and gas
inside the inner edge of the ring can thus remain optically thick in $^{13}$CO, but
be undetectable in dust.

The Gaussians at $\sim$250 and $\sim$370~AU trace features at the outer radius and beyond the dense, dust ring.
Instead of being interpreted in terms of bumps, these radial profiles can be equally well interpreted
as resulting from the presence of two ``dips'' or unresolved gaps at radii $\sim 250$ and $\sim 350$ AU
on top of a monotonic decrease with radius.

\subsection{The gas dynamics traced by CO}\label{ana:co_dyn}

\begin{figure*}[ht!]
\includegraphics[width=16cm]{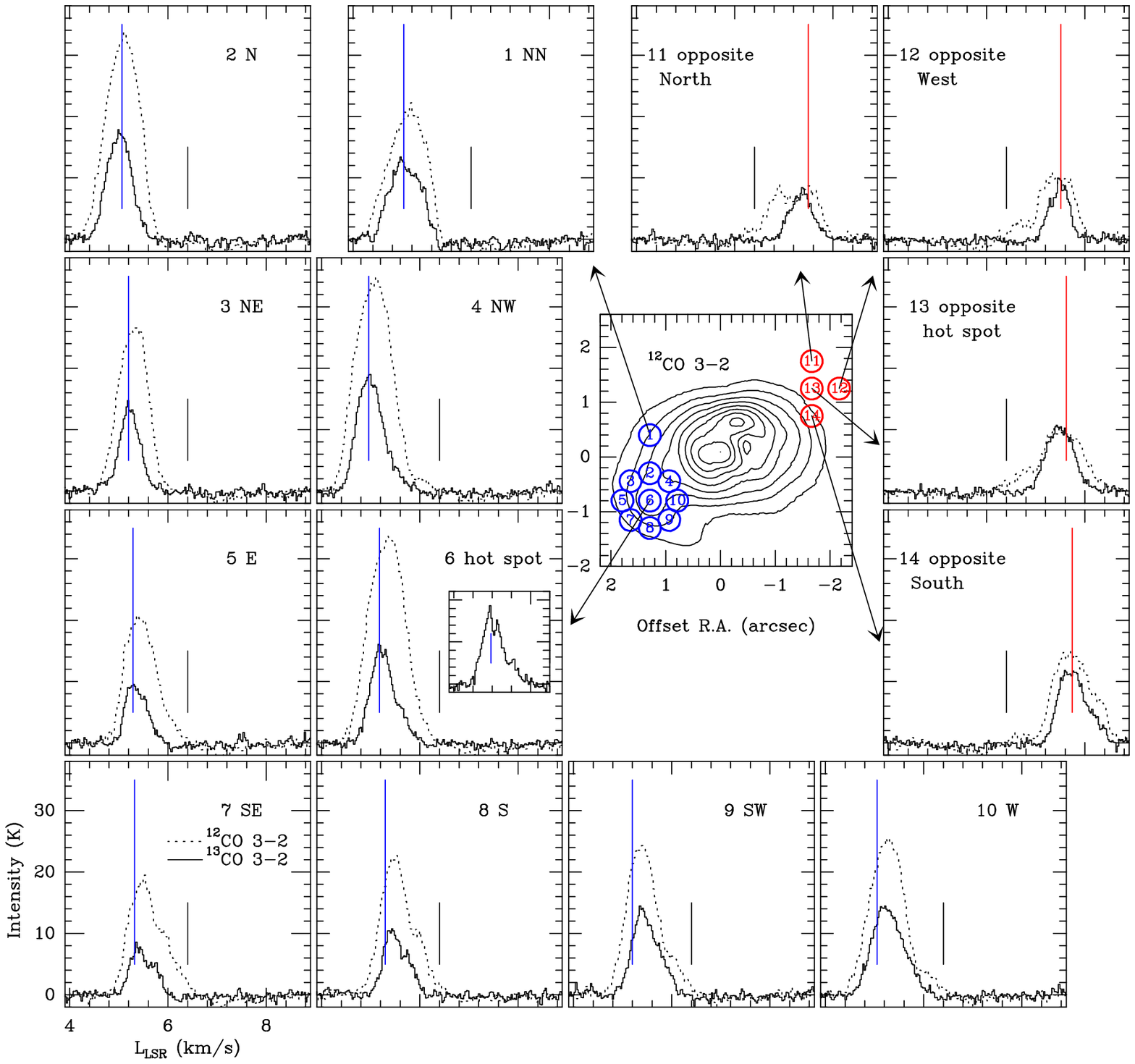}
\caption{Spectra of $^{12}$CO 3-2 (dashed) and $^{13}$CO 3-2 (solid) taken near the hot spot and at the opposite of the hot spot with respect to the disk center, obtained in natural weighting.
The intensity uncertainty in natural weighting is 0.4 K and 0.5 K for $^{12}$CO and $^{13}$CO, respectively.
The blue and red circles mark the locations where the spectra are taken.
The inset spectrum at the hot spot location
is the $^{13}$CO 3-2 line with robust weighting, displayed with a V$_{\rm LSR}$ scale from 4.3 to 6.4\,km/s
and an intensity scale from -2 to 22\,K. The blue and red segments mark the Keplerian velocities at
corresponding locations. The black segments mark the V$_{\rm sys}$ of 6.4\,km/s.}
\label{fig:spec}
\end{figure*}

In Figure \ref{fig:spec}, we present spectra of $^{12}$CO and
$^{13}$CO 3-2 lines extracted from the images with natural weighting
near the hot spot
 and at positions on the western side of the disk opposite the hot spot.
As all the images were self-calibrated,
deconvolution artefacts should be negligible here.
Even if most spectra are more complex than a simple Gaussian, some general trends
can be found.
On the western side, the spectra are mostly centered around the
expected Keplerian velocity, as indicated by the vertical red lines.
The $^{13}$CO spectra essentially show a single smooth component, while
the $^{12}$CO spectra are broader, with some
blue or red shifted tails or even double-peaked.
Double-peaked profiles can appear when CO
is located only at the upper layers of the gas disk, as in the case of
the CO disk orbiting the single star HD163296 \citep{deGregorio-Monsalvo13}.
This happens because the two layers of CO gas (above and below the
mid-plane) do not project identically.

Around the hot spot on the eastern side of the disk,
the $^{12}$CO and $^{13}$CO 3-2 spectra exhibit a component
at the Keplerian velocity (indicated by the vertical blue lines)
and a red-shifted tail (at velocity 5.7 km/s), which traces receding gas along the line of sight.
On the hot spot itself, the  $^{13}$CO spectrum obtained in robust weighting, although
noisier, further suggests a double-peaked
profile with separation of 0.16 km/s centered around the local Keplerian velocity.
In addition, the red-shifted component
is more intense on the hot spot, suggesting an intrinsic origin  at the location of the hot spot.

\subsection{Excitation conditions}\label{ana:exc_c}

We also use locations 6 and 12 to 14  shown in Figure \ref{fig:spec} to derive the
excitation conditions from the $^{12}$CO and $^{13}$CO 3-2 intensities.

On the western side, both lines have roughly the same intensity.
As the $^{13}$CO molecule is 70 times less abundant than the $^{12}$CO molecule,
the former should trace emission closer to the mid-plane of the disk if both lines are optically thick.
The similar intensity suggests optically thick emission from material at about
the same kinetic temperature \citep[see][for details about line formation in disks]{dartois03}.
This can only happen if the vertical temperature gradient is weak.
Such a small gradient can be expected, if most of the stellar light is
intercepted by the puffed up rim located at the ring inner edge (at radius 180-190 AU).
The outer disk  ($r > 250-300$ AU) then remains in  shadow and is thus relatively cool.

The H$_{2}$ density within the ring and outer disk is estimated to be large, being 10$^6$ to 10$^9$ cm$^{-3}$ \citep{guilloteau99}.
At such densities, the $^{12}$CO 3-2 and $^{13}$CO 3-2 are necessarily thermalized.
The brightness temperature is given by
$$ T_b = f (1-\exp{(-\tau)}) (J_\nu(T_\mathrm{ex}) - J_\nu(T_\mathrm{bg})$$
where $J_\nu(T)$ is the radiation temperature
\citep[$J_\nu(T) \approx T + h\nu/(2 k)$ for temperatures $T>h\nu/k$, see e.g.][]{Wilson2013},
$f$ is the beam filling factor
and $\tau$ the line opacity. As the lines are optically thick,
the brightness temperatures shown in Figure \ref{fig:spec} provide a direct measurement of the kinetic temperature \citep{dutrey07}.
The gas temperature at locations 12 to 14 is then 16-19 K with typical formal errors of 1 K, consistent with
the value derived from $^{13}$CO 2-1 by \citet{guilloteau99}.
We note that a steep temperature gradient with exponent $q=1$, i.e. temperature falling almost as 1/r, was found
by \citet{guilloteau99} in the ring and outer disk.
A similar gradient was reported for dust by  \citet{dutrey14}. Such a steep gradient is very different from those
found in disks without cavities where $q\approx 0.5$ \citep[e.g.][]{pietu07}.
This further supports the hypothesis that the inner edge of the ring is directly exposed to the stellar light,
casting a shadow on the outer disk.

We further estimate the column density using
RADEX\footnote{http://home.strw.leidenuniv.nl/$\sim$moldata/radex.php} \citep{van07}.
At the locations 12 to 14 in Figure \ref{fig:spec},
the column density N($^{13}$CO) is $>$3$\times$10$^{15}$cm$^{-2}$, assuming a line width of 0.3 km/s
and a temperature of 20 K.
This N($^{13}$CO) is  consistent with the analysis from \citet{guilloteau99}.

At the hot spot location (location 6), the brightness temperature
of $^{12}$CO 3-2, $\sim 36$\,K,
is identical to that found for the 2-1 line by \citet{dutrey14}, and much higher than
that of  $^{13}$CO 3-2, $\sim 16$\,K.
The same analysis as above implies a kinetic temperature of at least 42 K for $^{12}$CO.
Further assuming $^{12}$CO and $^{13}$CO emission
arise from the same medium and an isotologue ratio of 70, we can derive
the opacity, and thus refine the temperature estimate and determine the column density.
Under this assumption and with a beam filling factor $f=1$, the kinetic temperature is $45$\,K,
and N($^{13}$CO) $\approx 2\times10^{15}$cm$^{-2}$. These values would be larger for $f < 1$.
Given the high signal-to-noise ratio on the spectra, statistical errors are here
well below 1 K. Thus the ``hot spot'' is warmer than the disk by at least 20 K, with
a significance level well above 10 $\sigma$.

\section{Discussion}
\subsection{Disk gaps beyond the dense ring?}\label{sec:rings}

The presence of several dips (see section \ref{sec:ana:radial}) suggests that the observed CO radial
distributions are more complex than a smooth radial law.
The comparison between $^{12}$CO and $^{13}$CO indicates that
this brightness distribution seen around GG Tau A reflects the kinetic temperature (or more precisely
the product of temperature and beam filling factors) but
not directly the molecular column densities, because both lines are optically thick.
We discuss the possible origin of the shape of the radial brightness distribution below.

It is unlikely that the temperature difference between the rings
is driven by differences in
the illumination of the disk by the central stars.
As discussed in Sect.\ref{ana:exc_c}, most of the stellar light is
probably intercepted at the inner edge of the dense ring.
Indeed, the innermost CO ring at 170-176 AU precisely corresponds to this heated rim.
At larger distances, the lack of illumination results in low vertical temperature gradients,
because the disk is in the shadow of the ring.
It is possible that differences in surface density can cause apparent temperature variations.
With larger column densities, $^{12}$CO would sample higher layers in the disk and thus higher temperatures.
However, this effect should be smaller for $^{13}$CO because of its smaller abundance.
One way to produce the observed contrast in $^{12}$CO and $^{13}$CO radial distributions would
be a contrast in beam filling factors, as would be expected for a series of (radially) unresolved
gaps and dense annuli (which remain optically thick even in $^{13}$CO).
Unresolved spiral patterns can also produce a similar result.
The beam size (resolution) of our observation is $\sim$50 AU. If the annuli/spirals are not well
separated, this would cause an dip/drop in brightness at the location of gaps.

Such configurations are reminiscent of what is observed around  HD100546 \citep{pineda14} or  AB Auriga \citep{tang12}.
This is also expected in case of gravitational disturbances by an embedded object located in the
 disk \citep{pierens+13}. A distribution similar to that observed in the HL Tau dust disk \citep{brogan+15},
 scaled to the larger size of the GG Tau ring, can also mimic what we observe here.

We note that 80\% of the dust mass of the outer disk is confined at radii between 190 and 260 AU \citep{dutrey94}.
The gas rings may act like a dust trap
\citep{pinilla15,andrews14}. However, the model of \citet{andrews14} requires
the gas to extend down to 100 AU, while the inner edge seen in $^{13}$CO is at 170 AU.
A more comprehensive model is needed to see if another configuration could provide a viable solution.

\subsection{The CO hot spot and its surroundings}

We find a significantly higher temperature at the hot spot
location (by $>$ 20 K) as compared to the other locations at the same radius.
It seems unlikely that such a temperature enhancement is created
only by variations of the stellar illumination induced by homogeneities
in the disk. All mm continuum maps \citep{dutrey14,andrews14} and near-IR images  \citep{itoh14}
of GG Tau A show a very homogenous distribution of dust as a function of azimuth.
Although \citet{silber00} reported  a distortion in the near-IR emission
near the location of the hot spot in the HST image of polarized scattered
light (shown in Figure \ref{fig:gas_maps}), this was not confirmed in more recent
images (see discussion in \citeauthor{duchene04} 2004, and also \citeauthor{itoh14} 2014).
In addition, the strongest streamer seen in CO 3-2 and 6-5 would attenuate
the stellar light between GG Tau A and the hot spot, because it  appears at the same azimuth.
Finally, the CO gas temperature at 260 AU in the disks around the much more luminous Herbig Ae stars MWC\,480
and AB\,Aur is only 26 K and 32 K respectively \citep{pietu07,pietu05}.
Given these considerations, we find variations in stellar illumination highly unlikely to be the
cause in the temperature enhancement observed for the ``gas hot spot'' feature.

The observed increase in temperature and $^{12}$CO emission requires either a significant increase in the local
surface density as described above for the ring-like structures and/or a source of heating that affects a localized
region of the disk.  One such plausible explanation may be tidal disturbances induced by GG~Tau~B, a very low-mass
binary T Tauri system with mass of~0.17~M$_\odot$.  However, \citet{beust06} investigated such effects and found
that the induced disk distortions should affect a substantially large area of the disk, much larger than the
size of the hot spot feature.

Another plausible explanation is that the localized heating and density enhancements are provided by the
formation of an embedded substellar object \citep{gressel13}, similar to that imaged by \citet{delorme13} at a distance
of 80~AU from the the low-mass binary 2MASS J01033563-5515561(AB).  While recognizing the speculative nature of this
possibility, the formation of a companion near the outer edge of the dense and massive circum-ternary ring provides a
reasonable explanation for the existence and nature of the ring.  In addition, such an object would be expected to
clear a gap within the disk, which is consistent with the location of one of the dips suggested by the $^{13}$CO and
$^{12}$CO radial profile (see Section \ref{sec:ana:radial}).
We note that a planetary mass object at this distance
would not necessarily generate an open gap. A fully opened gap can be created only if the planet Hill's radius is larger
than the scale height, typically of the order of 30 AU there, and if the viscosity is small.
A planetary companion would most likely produce a surface density contrast instead of a complete disk gap \citep{Crida06}.

The observed distortion of the channel pattern seen in Figure \ref{fig:channels} is also in
qualitative agreement with the simulations of CO observations with ALMA of an embedded planet
in a Keplerian disk \citep{perez15}.
A double-peaked profile is detected (though marginally) at the location of the hot spot in $^{13}$CO 3-2.
If confirmed, this might be the signature of the circumplanetary (CP) disk \citep{perez15}.
The separation of the $^{13}$CO 3-2 peaks is 0.16\,km/s. Assuming the CP disk inclination is
that of the whole system (i.e. 36.3$\degr$), we can get a rough estimate of the mass of the embedded object.
Its mass is $\sim$0.16 Jupiter mass ($M_{\rm J}$), for an outer disk radius, r$_{\rm out}$, of 10 AU
or 0.016 $M_{\rm J}$ for r$_{\rm out}$=1 AU. On the other hand, the mass of the CO hot spot,
derived from the column density of $^{13}$CO, is much larger, $\sim2$ $M_{\rm J}$.
This indicates that, within a radial extension of 50 AU (our spatial resolution), a large amount of material is
still surrounding the putative embedded object.
In this scenario, the receding gas (i.e. the redshifted tail with respect to the local Keplerian velocity) may partly trace material located
in the vortex above the disk mid-plane and falling down onto the mid-plane on the proto-planet,
as reported by \cite{gressel13} in their simulations.

\section{Summary}

We report our new results for the GG Tau A system using ALMA in $^{12}$CO and $^{13}$CO 3-2
lines and in continuum at 0.9 mm.  The $^{12}$CO gas is found in the cavity of the dust ring and in the outer disk, while $^{13}$CO gas is detected only in the outer disk. At the outer edge of the dust ring, the hot spot is observed
both in $^{12}$CO and $^{13}$CO 3-2. In the outer disk, the radial distribution of
the brightness of the two CO isotopologues is complex and exhibits a series of ring-like features.
We note that the first dip coincides with the location of the hot spot, 250-260 AU from the disk center.
The angular resolution of these data does not allow a detailed analysis of such structures.
Studies of the excitation conditions beyond the ring, in the outer disk,  indicate that the
outer disk remains in the shadow of the ring.
Therefore,  the most probable origin of the hot spot is a local source of heating.
Together with the apparent dips,
this further supports our scenario of a young planet accreting
material from its surroundings.
Using ALMA, the achievable line sensitivity in a reasonable amount of time
still allows an improvement of the angular resolution by a factor 3 (0$\farcs$1).
With such angular resolution, analysis of the close surroundings of accreting planets and
quantitative comparison with model predictions may become feasible.

{\it Facilities:} \facility{PdBI}, \facility{ALMA}.
\acknowledgments{We acknowledge the anonymous referee for the comments which helped to make the manuscript clearer.
This paper makes use of the following ALMA data: ADS/JAO.ALMA\#2012.1.00129.S . ALMA is a partnership of ESO
(representing its member states), NSF (USA) and NINS (Japan), together with NRC (Canada), NSC and ASIAA (Taiwan), and
KASI (Republic of Korea), in cooperation with the Republic of Chile. The Joint ALMA Observatory is operated by ESO,
auI/NRAO and NAOJ. Y. T. acknowledges Tim Gledhill for providing the FITS file of the HST NICMOS image. This research
was partially supported by MOST grant MOST 103-2119-M-001-010-MY2. This research was partially supported by PNPS, the
French national program for stellar physics.}
\bibliographystyle{apj}                       

\end{document}